\documentclass[a4paper]{jpconf}
\usepackage{graphicx}
\usepackage{threeparttable}
\begin{document}
\title{Long-lived energetic particle source regions on the Sun}

\author{R Bu\v{c}\'ik$^{1, 2}$, D E Innes$^{1, 2}$, N H Chen$^1$, G M
Mason$^3$, R G\'omez-Herrero$^4$ and M E Wiedenbeck$^5$}

\address{$^1$ Max-Planck-Institut f\"{u}r Sonnensystemforschung, D-37077 G\"{o}ttingen, Germany}
\address{$^2$ Max Planck/Princeton Center for Plasma Physics, Princeton, NJ~08540, USA}
\address{$^3$ Applied Physics Laboratory, Johns Hopkins University, Laurel, MD~20723, USA}
\address{$^4$ Space Research Group, University of Alcal\'a, E-28871 Alcal\'a de Henares, Spain}
\address{$^5$ Jet Propulsion Laboratory, California Institute of Technology, Pasadena, CA~91109, USA}

\ead{bucik@mps.mpg.de}

\begin{abstract}
Discovered more than 40 years ago, impulsive solar energetic particle (SEP)
events are still poorly understood. The enormous abundance enhancement of the rare
$^{3}$He isotope is the most striking feature of these events, though large enhancements
in heavy and ultra-heavy nuclei are also observed. Recurrent $^{3}$He-rich SEPs in
impulsive events have only been observed for limited time periods, up to a few
days which is typically the time that a single stationary spacecraft is
magnetically connected to the source active regions on the Sun. With the launch
of the two STEREO spacecraft we now have the possibility of longer connection
time to solar active regions. We examined the evolution
of source regions showing repeated $^{3}$He-rich SEP emissions for
relatively long time periods. We found that recurrent $^{3}$He-rich SEPs in
these long-lived sources occur after the emergence of magnetic flux.
\end{abstract}

\section{Introduction}
The acceleration of $^{3}$He-rich solar energetic particles (SEPs) and their escape into
interplanetary space remains an unresolved question in solar physics. Anomalous
abundances with several orders of magnitude enhanced $^{3}$He ($\sim$10$^4$) and ultra-heavies ($\sim$10$^2$) suggest a unique
acceleration mechanism operating in the active regions on the Sun (see \cite{mas07}, for a review).

Recurrent $^{3}$He-rich SEP emissions observed by a single spacecraft during 1-2 days
\cite{rea86,mas99,mas00,wan06} or 2-3 days \cite{che15} have suggested a steadier production/release of energetic
ions from solar source regions \cite{pic06}. Multiple spacecraft
observations have recently shown that the source regions may produce recurrent SEPs for
a much longer time - about a quarter of a solar rotation \cite{buc14}.
It remains unclear what mechanisms in such long-lived sources lead to the repeated
particle emission. Is the production more stationary or rather intermittent
during such long periods?

To address these questions we examine in this paper a temporal evolution of the three
long-lived $^{3}$He-rich SEP source active regions (ARs). Two of them were previously reported (AR~11244, 11246)
and one is newly identified (AR~11045).

\begin{figure}
\begin{center}
\includegraphics[width=0.89\textwidth]{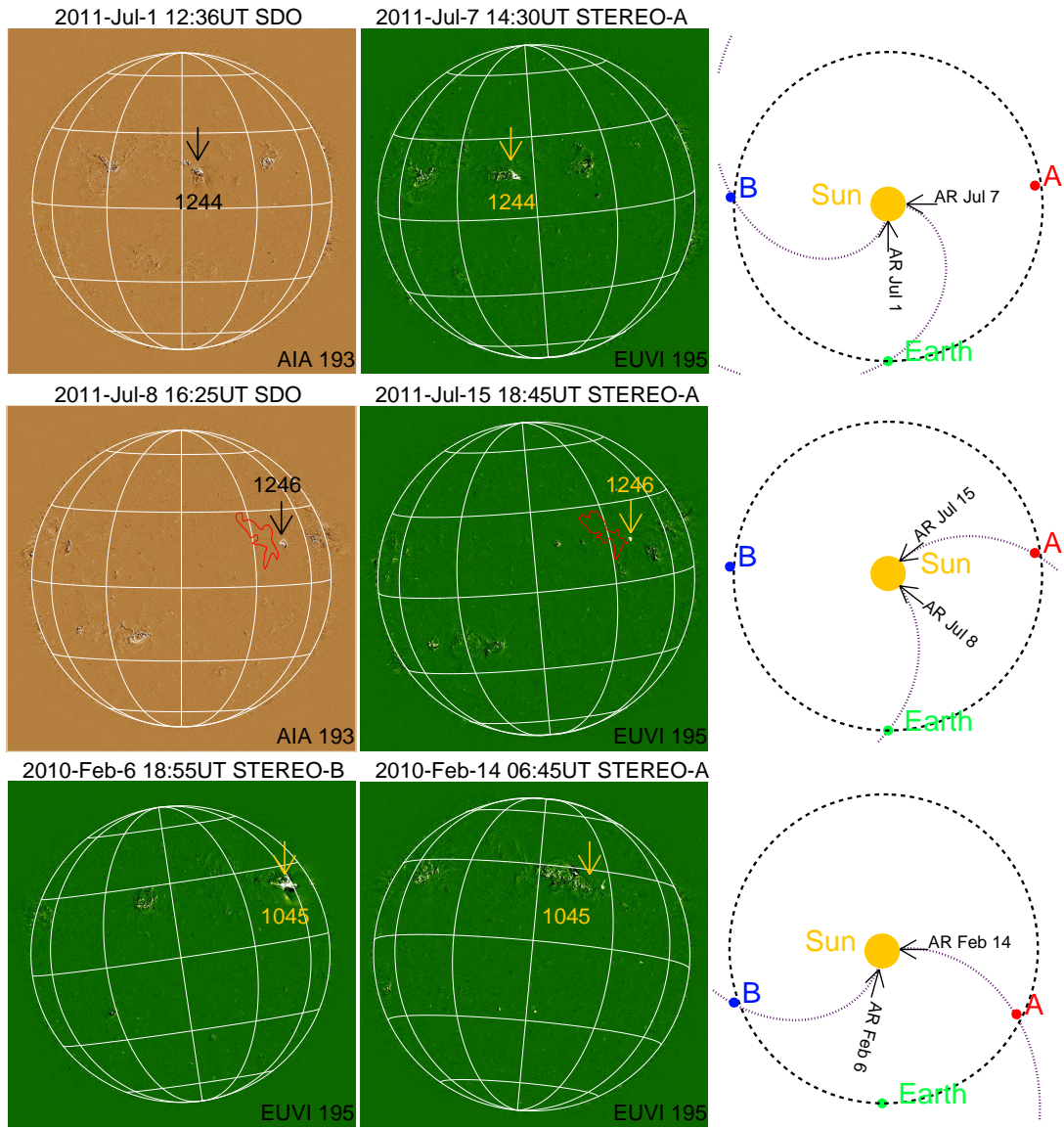}
\end{center}
\caption{\label{fig1}(Left, Middle) EUV 5 minute running difference images of the solar
disk around the times of the events associated type III radio bursts. Arrows
point to the long-lived $^{3}$He-rich SEP source ARs. Red contours indicate approximate
coronal hole boundaries as seen in direct EUV images. (Right) Ecliptic positions of
STEREO-A and -B spacecraft during the events. Parker spirals are shown to illustrate the spacecraft
magnetic connection to the solar sources.}
\end{figure}

\section{Observations}
An overview of the source ARs' locations on the solar disk and the spacecraft
ecliptic positions during the corresponding $^{3}$He-rich SEP events is shown in
Figure~\ref{fig1}. Left and middle columns of Figure~\ref{fig1} show AIA \cite{lem12}
and EUVI \cite{how08} extreme ultraviolet (EUV) running difference
images of the long-lived $^{3}$He-rich SEP source ARs~11244,
11246 and 11045 close to the time of events associated type III radio burst onsets.
The events from ARs~11244 and 11246 were examined in detail in \cite{buc14}.
AR~11244 was responsible for the 2011 July 1 $^{3}$He-rich SEP event observed by STEREO-B
and the July 7 event observed at the Earth (L1) by ACE. In both cases AR~11244 was near the west limb,
at W97 or W87. Figure~\ref{fig1} (upper row) shows this AR near the central meridian
from the Earth (SDO) and STEREO-A views. AR~11246 was responsible for the 2011 July 9 event
on ACE and the July 16 event on STEREO-A. This source was located at the coronal-hole
boundary ($\sim$W45) and had a quite small size. In contrast, AR~11045 was quite
sizeable and was the source of the 2010 February 6 and 7 events on STEREO-B,
the February 8 event on ACE \cite{wie13} and the February 14 event on
STEREO-A (see \ref{App}). The $^{3}$He increase on February 6
contained too much spillover from $^{4}$He to be unambiguously identified as
a $^{3}$He-rich event \cite{wie13}. However, SIT \cite{mas08} on STEREO-B shows low energy
Fe/O enhanced to $\sim$1 indicating a possible impulsive flare origin of this $^{3}$He
increase. Table~\ref{tab1} lists $^{3}$He-rich SEP events and sources
examined in this paper.

\begin{table}
\lineup
\begin{center}
\begin{threeparttable}
\caption{$^{3}$He-rich SEP events \label{tab1}}
\begin{tabular}{lclcl}
\br
Start time\tnote{a} & Spacecraft & Type III burst & AR & AR location\tnote{b}\\
\mr
2011 Jul \m1.9 & B & Jul \m1 12:36 & 11244 & N14W06\\
2011 Jul \m7.9 & ACE & Jul \m7 14:31 & 11244 & N14W87\\
2011 Jul \m9.0 & ACE & Jul \m8 16:25 & 11246 & N15W45\\
2011 Jul \;16.4 & A & Jul \;15 18:46 & 11246 & N15W138\\
2010 Feb \07.0 & B & Feb \06 18:53 & 11045 & N21E15\\
2010 Feb 14.6 & A & Feb 14 06:45 & 11045 & N21W91\\
\br
\end{tabular}
\begin{tablenotes}
\item[a] at 275~keV\,nucleon$^{-1}$; nominal travel time of ions of this energy from Sun to 1~AU is about 0.3
day
\item[b] from the Earth view
\end{tablenotes}
\end{threeparttable}
\end{center}
\end{table}

\begin{figure}
\begin{center}
\includegraphics[width=0.8\textwidth]{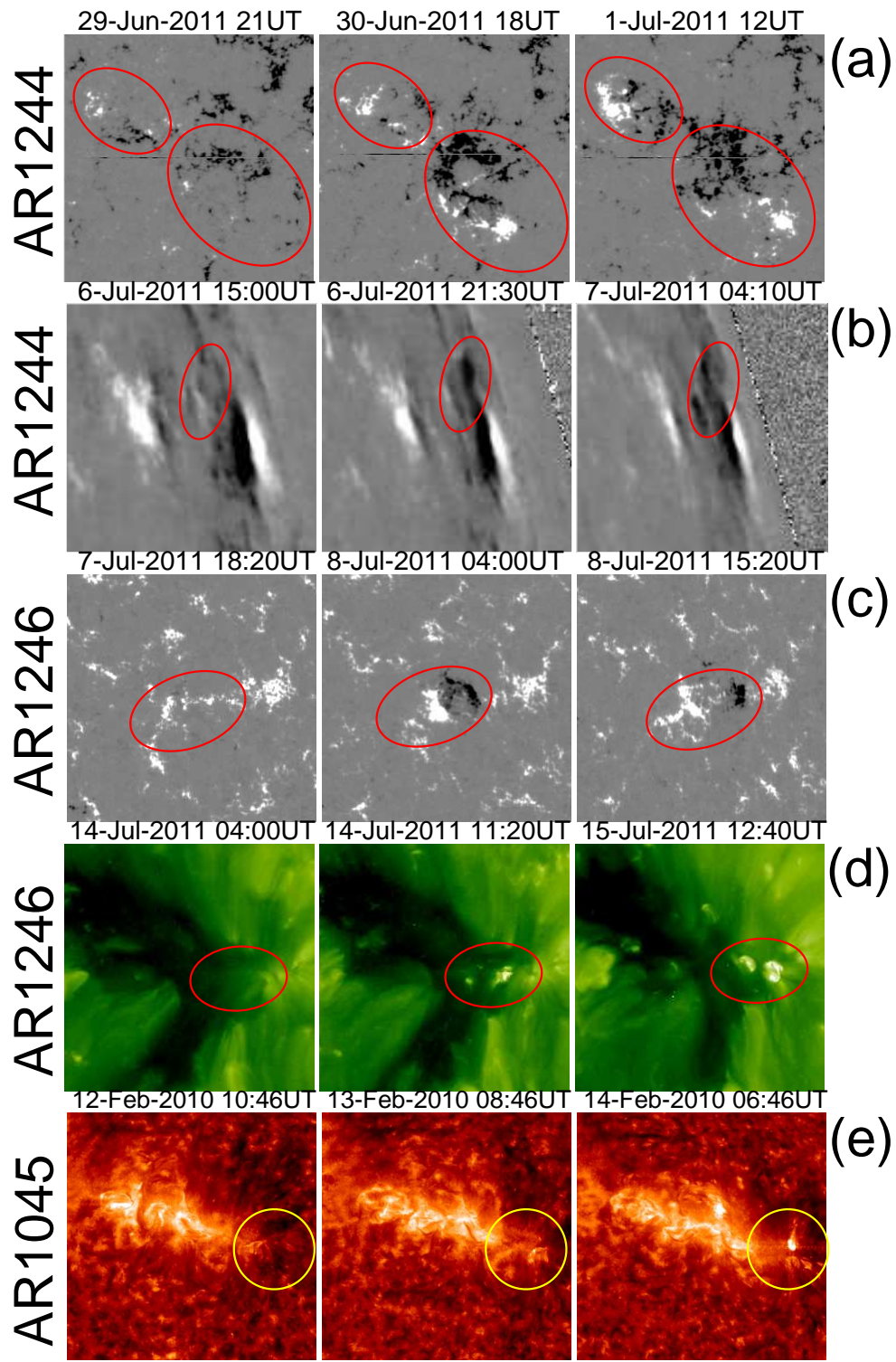}
\end{center}
\caption{\label{fig2}(a, b, c) SDO HMI magnetograms. (d) STEREO-A EUVI 195~{\AA}
images. (e) STEREO-A EUVI 304~{\AA} images. The ellipses mark the areas of the magnetic
flux emergence (a, b, c) or new flaring activity (d, e).}
\end{figure}

Figure~\ref{fig2} shows the evolution of the source ARs in SDO HMI \cite{sch12}
magnetograms (2011 July 1, July 7 and July 9 events) or in STEREO-A EUVI
images (2011 July 16, 2010 February 14 events). For each
event a sequence of three images is shown. The first image is around the beginning
of the region emergence and the third is around the particle injection time
from that region. The magnetic field evolution in AR~11045 (the source of the 2010
February 6 event) has been examined in detail in \cite{li12}.

AR~11244 had started to form by the emergence of a positive polarity magnetic
flux in a negative polarity area at the end of 2011 June 29 (see Figure~\ref{fig2}a),
$\sim$1.5~day before the ion injection in the July 1 $^{3}$He-rich SEP event.
The potential-field source-surface (PFSS) extrapolations indicate that open field in AR~11244 has negative
polarity \cite{buc14}. Another magnetic flux emergence (of negative
polarity) occurred in the same AR~11244 on 2011 July 6 (see Figure~\ref{fig2}b), $\sim$1 day
before the $^{3}$He-rich SEP injection. The event was accompanied
by multiple electron events with the first type III burst at 05:10~UT.

The creation of AR~11246 was associated with a negative polarity flux emergence
in a coronal hole (of a positive polarity) on 2011 July 7 (see Figure~\ref{fig2}c), $\sim$1~day
before the ion injection. The emergence covered a small and compact
region in contrast to AR~11244. In 2011 July 16 event, the source AR~11246 was out of
the Earth view and therefore its evolution could be seen only in STEREO-A EUV
images. Figure~\ref{fig2}d shows a new brightening at the AR 11246 location on
July 14 which may be related to further magnetic flux emergence. This occurred $\sim$1 day before the SEP injection. Note
that the event was associated with several type III bursts starting at 12:40~UT
on July 15.

AR~11045 emerged on 2010 February 5 with a positive polarity field into a negative
polarity area \cite{li12}, $\sim$1.5 day before the ion injection observed
by STEREO-B. The emerging phase of this AR ended on February 8. The PFSS
extrapolations along with interplanetary magnetic field (IMF) polarity suggest that ACE was connected
to AR~11045 for several days at least until
the AR rotated behind the limb on February 14 but no further event on ACE was observed.
Li et al. \cite{li12} report another magnetic flux increase in AR~11045
on February 12 when the region was close to the west limb. STEREO-A EUV images
(see Figure~\ref{fig2}e) show a creation of a new flaring area in the west edge of AR~11045,
perhaps related to the same magnetic flux increase. This emergence occurred $\sim$1.5~day
before $^{3}$He-rich SEP injection observed by STEREO-A.

\section{Summary and discussion}

We investigated the temporal evolution of previously reported long-lasting $^{3}$He-rich
SEP source ARs~11244 and 11246. We also identified a new long-lasting source
AR 11045 and examined its evolution. We found that recurrent $^{3}$He-rich SEPs in
the long-lived sources occur after the emergence of magnetic flux. All these
ARs were newly emerging with the first particle emission occurring within $\sim$1-2 days
after the emergence. Interestingly, the next $^{3}$He-rich SEP event in the same source
was associated with additional magnetic flux emergence. The EUV images or PFSS
extrapolations indicate that all three sources probably emerged into pre-existing
open field.

Our observations are consistent with an earlier example of particle injection
in the 2002 December 12 $^{3}$He-rich SEP event, which occurred about 1 day after the emergence
of magnetic flux in a unipolar area \cite{wan06}. The authors have suggested
that such a process may lead to magnetic reconnection, followed by particle
acceleration on open field lines.

The presented long-lasting sources exhibited a wide range of coronal activities.
For example, AR~11244 produced several cool, surge-like eruptions, as well as
a number of GOES B-class X-ray flares. In contrast, AR~11246 produced
only small EUV brightenings. AR~11045 was the most active region with several
B-, C- and M-class flares (as well as hard X-rays) and few slow coronal mass ejections. This
likely implies that there is no preferred type of $^{3}$He-rich SEP source.
It appears that any AR can be a $^{3}$He-rich SEP source if there is magnetic flux emergence
near the open field region.

\ack
This work was supported by the Max-Planck-Gesellschaft zur F\"{o}rderung der Wissenschaften.
The STEREO SIT is supported by the Bundesministerium f\"{u}r Wirtschaft through the Deutsches
Zentrum f\"{u}r Luft- und Raumfahrt (DLR) under grant 50 OC 1301. ACE/ULEIS and
STEREO/SIT are supported at APL by NASA grant NNX13AR20G/115828 and NASA through
subcontract SA4889-26309 from the University of California Berkeley. R.B. thanks Gary Zank for
inviting to the 14th AIAC that led to this paper.

\appendix
\section{}\label{App}
\begin{figure}
\begin{center}
\includegraphics[width=0.9\textwidth]{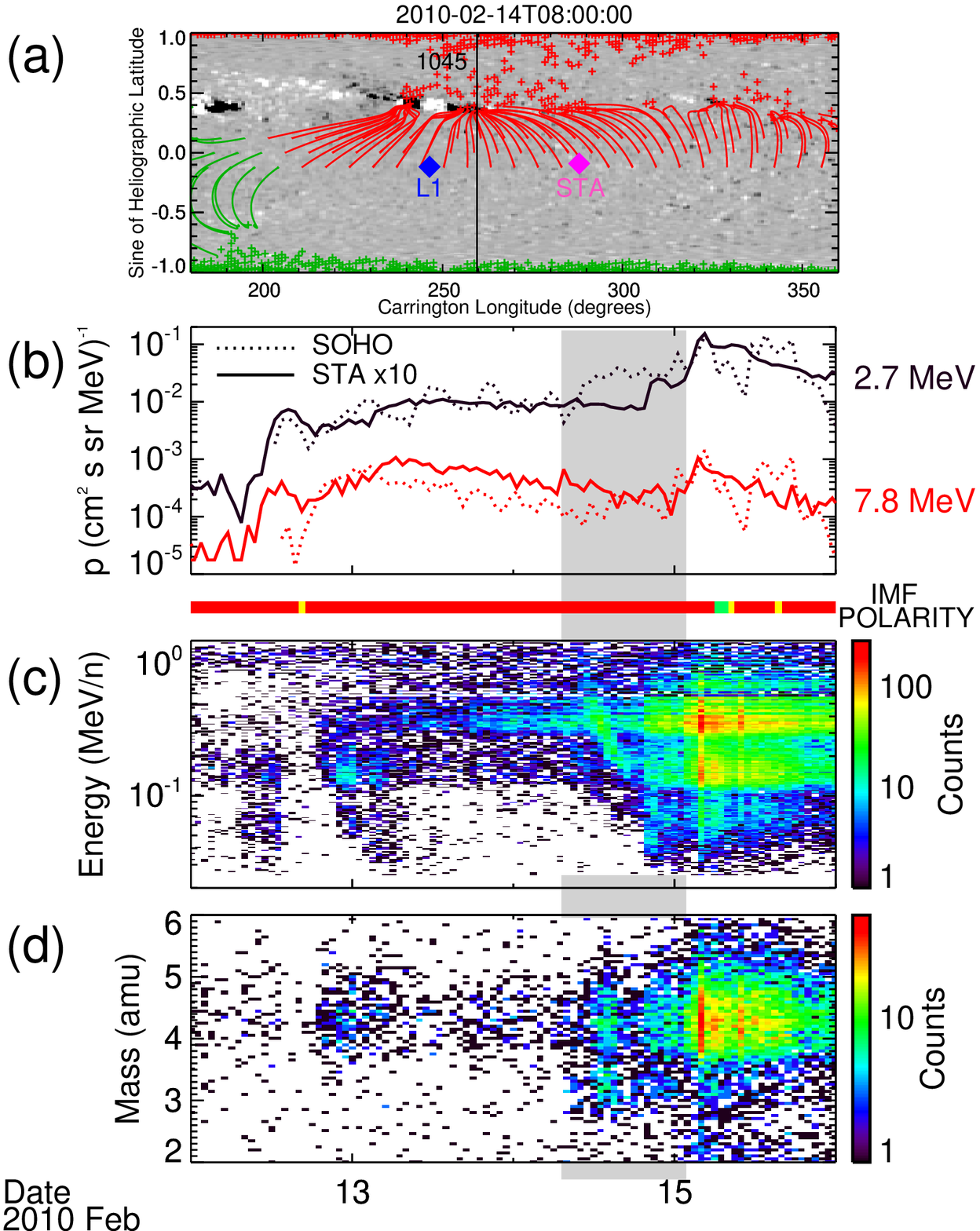}
\end{center}
\caption{\label{figA1}(a) Photospheric magnetic field with PFSS model coronal
field lines (red - negative and green - positive polarity) around the injection
time in STEREO-A February 14 $^{3}$He-rich SEP event. Shown are field lines
which intersect source surface at latitudes 0$^{\circ}$ and $\pm$7$^{\circ}$. Diamonds mark L1 and
STEREO-A (STA) magnetic foot-points on the source surface. Black vertical line
marks west solar limb from the Earth view. (b) SOHO (L1) and STA proton intensities.
(c) SIT-A kinetic energy spectrogram of all ions. (d) SIT-A helium mass spectrogram
at 0.2-0.5~MeV\,nucleon$^{-1}$. Shaded region marks $^{3}$He-rich SEP event. }
\end{figure}

Figure~\ref{figA1} shows February 14 $^{3}$He-rich SEP event observed by STEREO-A. The spacecraft
was connected via negative polarity field lines to AR~11045 as indicated by the PFSS
extrapolations (Figure~\ref{figA1}a) along with IMF polarity. The event
shows velocity dispersion (Figure~\ref{figA1}c) and occurred during the large gradual
event observed by both SOHO and STEREO-A (Figure~\ref{figA1}b). Note that high ion counts in
SIT spectrograms (Figures~\ref{figA1}c,~\ref{figA1}d) after the $^{3}$He-rich
SEP event are due to interplanetary shock passage. Figure~\ref{figA1} indicates
that L1 spacecraft were also connected to AR~11045 but ACE did not observe the same
$^{3}$He injection. Perhaps ACE had a connection to different area in AR~11045.

\setcounter{section}{1}

\end{document}